\documentstyle[twoside,amsmath,amssymb,epsfig,verbatim]{article} 

\catcode`\@=11
\long\def\@makefntext#1{
\protect\noindent \hbox to 3.2pt {\hskip-.9pt  
$^{{\eightrm\@thefnmark}}$\hfil}#1\hfill}		
\def\thefootnote{\fnsymbol{footnote}}
\def\@makefnmark{\hbox to 0pt{$^{\@thefnmark}$\hss}}	
\def\ps@myheadings{\let\@mkboth\@gobbletwo
\def\@oddhead{\hbox{}
\rightmark\hfil\eightrm\thepage}   
\def\@oddfoot{}\def\@evenhead{\eightrm\thepage\hfil
\leftmark\hbox{}}\def\@evenfoot{}
\def\sectionmark##1{}\def\subsectionmark##1{}}

\oddsidemargin=\evensidemargin
\renewcommand{\thefootnote}{\fnsymbol{footnote}}

\newcounter{sectionc}\newcounter{subsectionc}\newcounter{subsubsectionc}
\renewcommand{\section}[1] {\vspace{12pt}\addtocounter{sectionc}{1} 
\setcounter{subsectionc}{0}\setcounter{subsubsectionc}{0}\noindent 
	{\tenbf\thesectionc. #1}\par\vspace{5pt}}
\renewcommand{\subsection}[1] {\vspace{12pt}\addtocounter{subsectionc}{1} 
	\setcounter{subsubsectionc}{0}\noindent 
	{\bf\thesectionc.\thesubsectionc. {\kern1pt \bfit #1}}\par\vspace{5pt}}
\renewcommand{\subsubsection}[1] {\vspace{12pt}\addtocounter{subsubsectionc}{1}
	\noindent{\tenrm\thesectionc.\thesubsectionc.\thesubsubsectionc.
	{\kern1pt \tenit #1}}\par\vspace{5pt}}

\newcounter{appendixc}
\newcounter{subappendixc}[appendixc]
\newcounter{subsubappendixc}[subappendixc]
\renewcommand{\thesubappendixc}{\Alph{appendixc}.\arabic{subappendixc}}
\renewcommand{\thesubsubappendixc}
	{\Alph{appendixc}.\arabic{subappendixc}.\arabic{subsubappendixc}}

\renewcommand{\appendix}[1] {\vspace{12pt}
        \refstepcounter{appendixc}
        \setcounter{figure}{0}
        \setcounter{table}{0}
        \setcounter{lemma}{0}
        \setcounter{theorem}{0}
        \setcounter{corollary}{0}
        \setcounter{definition}{0}
        \setcounter{equation}{0}
        \renewcommand{\thefigure}{\Alph{appendixc}.\arabic{figure}}
        \renewcommand{\thetable}{\Alph{appendixc}.\arabic{table}}
        \renewcommand{\theappendixc}{\Alph{appendixc}}
        \renewcommand{\thelemma}{\Alph{appendixc}.\arabic{lemma}}
        \renewcommand{\thetheorem}{\Alph{appendixc}.\arabic{theorem}}
        \renewcommand{\thedefinition}{\Alph{appendixc}.\arabic{definition}}
        \renewcommand{\thecorollary}{\Alph{appendixc}.\arabic{corollary}}
        \renewcommand{\theequation}{\Alph{appendixc}.\arabic{equation}}
        \noindent{\tenbf Appendix \theappendixc #1}\par\vspace{5pt}}
\newcommand{\subappendix}[1] {\vspace{12pt}
        \refstepcounter{subappendixc}
        \noindent{\bf Appendix \thesubappendixc. {\kern1pt \bfit #1}}
	\par\vspace{5pt}}
\newcommand{\subsubappendix}[1] {\vspace{12pt}
        \refstepcounter{subsubappendixc}
        \noindent{\rm Appendix \thesubsubappendixc. {\kern1pt \tenit #1}}
	\par\vspace{5pt}}

\newcommand{\textlineskip}{\baselineskip=13pt}
\newcommand{\smalllineskip}{\baselineskip=10pt}

\def\eightcirc{
\begin{picture}(0,0)
\put(4.4,1.8){\circle{6.5}}
\end{picture}}
\def\eightcopyright{\eightcirc\kern2.7pt\hbox{\eightrm c}} 

\newcommand{\copyrightheading}[1]
	{\vspace*{-2.5cm}\smalllineskip{\flushleft
	{\footnotesize International Journal of Modern Physics A, #1}\\
	{\footnotesize $\eightcopyright$\, World Scientific Publishing
	 Company}\\
	 }}

\def\abstracts#1#2#3{{
	\centering{\begin{minipage}{4.5in}\baselineskip=10pt\footnotesize
	\parindent=0pt #1\par 
	\parindent=15pt #2\par
	\parindent=15pt #3
	\end{minipage}}\par}} 

\newcommand{\bibit}{\nineit}

\renewenvironment{thebibliography}[1]
	{\frenchspacing
	 \ninerm\baselineskip=11pt
	 \begin{list}{\arabic{enumi}.}
	{\usecounter{enumi}\setlength{\parsep}{0pt}
	 \setlength{\leftmargin 12.7pt}{\rightmargin 0pt} 
	 \setlength{\itemsep}{0pt} \settowidth
	{\labelwidth}{#1.}\sloppy}}{\end{list}}

\newcounter{itemlistc}
\newcounter{romanlistc}
\newcounter{alphlistc}
\newcounter{arabiclistc}

\newcommand{\fcaption}[1]{
        \refstepcounter{figure}
        \setbox\@tempboxa = \hbox{\footnotesize Fig.~\thefigure. #1}
        \ifdim \wd\@tempboxa > 5in
           {\begin{center}
        \parbox{5in}{\footnotesize\smalllineskip Fig.~\thefigure. #1}
            \end{center}}
        \else
             {\begin{center}
             {\footnotesize Fig.~\thefigure. #1}
              \end{center}}
        \fi}

\newcommand{\tcaption}[1]{
        \refstepcounter{table}
        \setbox\@tempboxa = \hbox{\footnotesize Table~\thetable. #1}
        \ifdim \wd\@tempboxa > 5in
           {\begin{center}
        \parbox{5in}{\footnotesize\smalllineskip Table~\thetable. #1}
            \end{center}}
        \else
             {\begin{center}
             {\footnotesize Table~\thetable. #1}
              \end{center}}
        \fi}

\def\@citex[#1]#2{\if@filesw\immediate\write\@auxout
	{\string\citation{#2}}\fi
\def\@citea{}\@cite{\@for\@citeb:=#2\do
	{\@citea\def\@citea{,}\@ifundefined
	{b@\@citeb}{{\bf ?}\@warning
	{Citation `\@citeb' on page \thepage \space undefined}}
	{\csname b@\@citeb\endcsname}}}{#1}}

\newif\if@cghi
\def\cite{\@cghitrue\@ifnextchar [{\@tempswatrue
	\@citex}{\@tempswafalse\@citex[]}}
\def\citelow{\@cghifalse\@ifnextchar [{\@tempswatrue
	\@citex}{\@tempswafalse\@citex[]}}
\def\@cite#1#2{{$\null^{#1}$\if@tempswa\typeout
	{IJCGA warning: optional citation argument 
	ignored: `#2'} \fi}}

\def\pmb#1{\setbox0=\hbox{#1}
	\kern-.025em\copy0\kern-\wd0
	\kern.05em\copy0\kern-\wd0
	\kern-.025em\raise.0433em\box0}

\def\fnt#1#2{\footnotetext{\kern-.3em
	{$^{\mbox{\scriptsize #1}}$}{#2}}}

\def\fpage#1{\begingroup
\voffset=.3in
\thispagestyle{empty}\begin{table}[b]\centerline{\footnotesize #1}
	\end{table}\endgroup}

\def\runninghead#1#2{\pagestyle{myheadings}
\markboth{{\protect\footnotesize\it{\quad #1}}\hfill}
{\hfill{\protect\footnotesize\it{#2\quad}}}}
\font\tenrm=cmr10
\font\tenit=cmti10 
\font\tenbf=cmbx10
\font\bfit=cmbxti10 at 10pt
\font\ninerm=cmr9
\font\nineit=cmti9

\font\eightrm=cmr8

\def\qed{\hbox{${\vcenter{\vbox{			
   \hrule height 0.4pt\hbox{\vrule width 0.4pt height 6pt
   \kern5pt\vrule width 0.4pt}\hrule height 0.4pt}}}$}}

\renewcommand{\thefootnote}{\fnsymbol{footnote}}	

\newcommand{\be}{\begin{equation}}
\newcommand{\ee}{\end{equation}}

\newcommand{\formula}[2]
  { \begin{equation} \label{#1} #2 \end{equation} }

\begin{document}

\runninghead{Relaxing Near the Critical Point}
{Relaxing Near the Critical Point}

\normalsize\textlineskip
\thispagestyle{empty}
\setcounter{page}{1}

\copyrightheading{}			

\vspace*{0.88truein}

\fpage{1}
\centerline{\bf RELAXING NEAR THE CRITICAL POINT\footnote{Talk given
at DPF2000, August 9-12, 2000,  Columbus, Ohio.}}
\vspace*{0.37truein}
\centerline{\footnotesize M. SIMIONATO}
\vspace*{0.015truein}
\begin{center}{\footnotesize\it LPTHE, Universit\'e Pierre et Marie Curie 
(Paris VI) et Denis Diderot (Paris VII),\\ Tour 16, 1er \'etage,
4, Place Jussieu, 75252 Paris, Cedex 05, France\\ 
and\\ 
Istituto Nazionale di Fisica Nucleare, Rome, Italy\\
E-mail: micheles@lpthe.jussieu.fr}
\end{center}
\baselineskip=10pt
\vspace*{10pt}

\vspace*{0.21truein}
\abstracts{I present an analysis of the relaxation rate for 
long-wavelength fluctuations of the order parameter in an $O(N)$ scalar 
theory near the critical point. Our motivation is to model 
the non-equilibrium dynamics of
critical fluctuations near the chiral phase transition in QCD. In the 
next-to-leading order in the large $N$ expansion we find a critical slowing
down regime, i.e. an increasing of the relaxation time of long wavelengths
fluctuations. This result suggests, for near critical systems, relevant 
deviations from thermal equilibrium for the distribution functions of 
low-energy particles and could have important phenomenological consequences
in Heavy Ions Collision and
in the Early Universe Cosmology.}{}{}

\textheight=7.8truein
\setcounter{footnote}{0}
\renewcommand{\thefootnote}{\alph{footnote}}

\textlineskip			
\vspace*{12pt}			

\noindent
For QCD with only two flavors of massless quarks it has been
argued\cite{o4pis} that the chiral phase transition at finite
temperature is of second order and
described by the universality class of O(4) Heisenberg
ferromagnets. Second order critical points are characterized by
strong critical long-wavelength fluctuations and a diverging
correlation length that  could lead to 
important experimental signatures\cite{QCDPT}. These signatures would
be akin to  critical opalescence near  
the critical point in binary fluids and could be observed
in an event-by-event analysis of the fluctuations of the 
charged particle transverse momentum distribution (mainly pions)\cite{QCDPT}.

Critical 
slowing down of long-wavelength fluctuations near a second order
critical point is the statement that the 
long-wavelength Fourier components of the order parameter relax very
slowly towards equilibrium\cite{MA}. In mean-field 
theory in {\em classical critical phenomena},  the relaxation time for 
homogeneous fluctuations diverges as $\tau \propto \xi^z$ 
with  $\xi$ the
correlation length, or, at critical point, as $\tau(k) \propto k^z$ with $z$ a 
dynamical critical exponent\cite{MA}. 


The phenomenological importance of critical slowing down for the
QCD phase transition both in Relativistic Heavy Ion Collisions as well
as in Early Universe Cosmology motivates us to study this phenomenon
in a model Quantum Field Theory that bears on the low energy (chiral)
phenomenology of QCD, the $O(N)$ linear sigma model. 

We point out that 
the analysis requires non-perturbative techniques since
near the critical points perturbation theory in the quartic coupling $\lambda$
breaks down. 
This can be directly understood from a two-loop 
computation of the relaxation rate. We compute both the critical
damping rate $\Gamma(k,T_c)$ for inhomogenous fluctuations of wavevector
$\vec k$ and the near critical damping rate $ \Gamma_0(m_T,T)$ 
for homogenous fluctuations, where $ m_T \propto
|T-T_c|^{1/2}\ll T_c$ is the effective thermal mass.

We find\cite{CSD} respectively $\Gamma(k,T_c) \propto \lambda^2 T_c^2/k$
and $ \Gamma_0(m_T,T) \propto \lambda^2 T^2/m_T $.  These results hold
in the semisoft region $T\gg k\gg\lambda T$ and
clearly reveal the breakdown of the perturbative expansion in the long
wavelength limit $ k \rightarrow 0 $ at $ T = T_c $ and for $ T \to
T_c $ and $ k = 0 $. In order to face this problem analytically, we 
implement a non-perturbative
resummation of bubble-type diagrams via the large $ N $ approximation
and compute the damping rate in the next-to-leading order in the large $ N $
limit (alternatively, the use of the nonperturbative thermal 
renormalization group approach has been advocated by
various authors\cite{pietroni}). This resummation 
is akin to that obtained via the renormalization group with the 
one loop beta function and reveals the softening of the scattering
amplitude and the  crossover to an effective three dimensional theory
for momenta $q   \ll  \lambda T$.   

A detailed analysis of the different
contributions to the two-loops relaxation rate shows that the rate is 
dominated by very soft  loop momentum $q   \ll  \lambda T$ which in
the weak coupling limit $\lambda    \ll  1$ are classical. The
implementation of the non-perturbative large $ N $ resummation
 effectively entails a screening of
the contribution from these momenta. 
Consequently the most important contribution to the 
relaxation rate arises both from the {\em semisoft classical} region of loop
momentum $ T   \gg   q   \gg   \lambda T $ and also from the {\em hard}
region $ q\geq T $. A detailed analysis of the contribution from the
loop momenta reveals a non-perturbative {\em ultrasoft scale} 
$$ 
k_{us}\simeq{\lambda \; T \over 4 \pi }e^{-{4 \pi\over \lambda}} \; ,
$$ 
which is exponentially small in the small coupling regime.

We find\cite{CSD} that for soft momenta $ \lambda T\gg
k \gg k_{us} $ the damping rate is
dominated by classical semisoft loop momenta and given by 
\formula{G.soft}
{\Gamma(k,T_c) \buildrel{\lambda T\gg
k \gg k_{us}}\over= { \lambda T_c \over 2\pi N }
\left[ 1+ {\cal O}\left( {1 \over \ln{\lambda T_c \over k}}\right)\right]\; ,
}
i.e. it is essentially $k-$independent and depends linearly
on the coupling constant.
For $ k \ll k_{us} $ the classical approximation breaks down and the 
damping rate is given by
\formula{G.us}
{\Gamma(k,T_c) \buildrel{k \ll k_{us}}\over= { 4\,\pi \, T_c \over 3\,
N \ln{T_c\over k}}\left[1+{\cal O}\left({1\over \ln{T_c\over k}}\right)
\right] \; . 
}
For homogeneous  fluctuations near the critical point ($ k=0,~~m_T
\neq 0$) the damping rate is nearly constant
for $m_T\gg k_{us}$
\formula{G0.soft}
{\Gamma_0(m_T,T) \buildrel{m_T \gg k_{us}}\over= { \lambda T \over 2\pi N }
\left[ 1+ {\cal O}\left( {1 \over \ln{\lambda T \over {m_T}}}\right)\right]\; ,
}
whereas in the ultrasoft region is logarithmically vanishing
\formula{G0.us}
{ \Gamma_0(m_T,T) \buildrel{m_T \ll k_{us}}\over={ 4\,\pi \, T \over 3\,
N \ln{T\over m_T}}\left[1+{\cal O}\left({1\over \ln{T\over
m_T}}\right) \right]\; .  
}
Thus critical slowing down, i.e,  the vanishing of the quasiparticle width $
\Gamma $ for long-wavelengths  emerges in the ultrasoft limit $
k\ll k_{us}  $  or very near the critical point $ m_T \ll k_{us} $. 
We notice that in such regimes the rate is
independent of the coupling $ \lambda $.

The large $ N $ approximation is not limited to 
weak coupling and our results apply just as well to a strong coupling
case $ \lambda > 1$. In such a case we have $ k_{us}\sim \lambda T\gg k$, 
therefore the classical approximation is not
valid and the relaxation rate for $k\ll T$ or $m_T
\ll T$ is given by Eqs. (\ref{G.us}) and (\ref{G0.us}) respectively. 
The weak coupling analysis
instead clearly reveals that there 
emerges a {\em hierarchy} of widely separated scales for loop momenta: 
from hard $ q\geq T $ to semisoft $ T   \gg   q   \gg   \lambda T $, and
soft $ \lambda T   \gg   q $ that lead to different contributions to
the relaxation rate. Which is the relevant scale for the damping rate
is determined by the wavevector of the fluctuation of the order
parameter and the proximity to the critical temperature. 
For $ k~,~ m_T \gg k_{us} $ the classical 
approximation does apply
and the damping rate is dominated by the soft and semisoft classical
loop momenta [with the results (\ref{G.soft}) and (\ref{G0.soft})], 
whereas for $ k~,~m_T \ll
k_{us} $ the classical approximation breaks down and the damping rate
is dominated by hard loop momenta $ q\geq T $ [with the results
(\ref{G.us}) and (\ref{G0.us})].   

Further studies are needed in order to understand the effects of the 
next-to-next-to-leading corrections in $1/N$, which can be
be relevant for finite $N$ in the critical limits $k\to0$ at
$T=T_c$ and $m_T\to0$ at $k=0$.

\end{document}